\documentclass[11pt,a4paper]{article}
\usepackage{jcappub}
\usepackage{amsmath}
\usepackage{hyperref}
\hypersetup{colorlinks=True,citecolor=blue}
\usepackage{amsfonts}
\usepackage{amssymb}
\usepackage{caption}
\usepackage{subfigure}
\usepackage{float}
\usepackage{bm}
\usepackage{graphicx}
\usepackage{color}
\usepackage{verbatim}

\graphicspath{{}}

\title{Future evolution in a backreaction model and the analogous scalar field cosmology }
\author[\dagger]{Amna Ali,} 
\affiliation [\dagger]{S. N. Bose National Centre for Basic Sciences, Block JD, Sector-III, Salt Lake, Kolkata 700106, India.}
\author[\dagger]{A. S. Majumdar}

\abstract{We investigate the future evolution of the universe using the Buchert
framework for averaged backreaction in the context of a two-domain partition of the universe. We show that this approach allows for the possibility of 
the global acceleration vanishing at a finite future time, provided that none of the subdomains accelerate
individually. The model at large scales is analogously described in terms of a homogeneous scalar field emerging with a potential
that is fixed and free from phenomenological parametrization.
The dynamics of this scalar field is explored
in the analogous FLRW  cosmology. 
 We use observational data from Type Ia Supernovae, Baryon Acoustic Oscillations, and Cosmic Microwave Background to constrain the parameters of the model  for a viable cosmology,
providing the corresponding likelihood contours.}

\begin{document}

\maketitle

\section{Introduction}

The standard Friedmann-Lema\^{\i}tre-Robertson-Walker (FLRW) cosmological  model relies on the Cosmological Principle \textit{i.e.}, the universe 
is exactly homogeneous and isotropic. The Cosmological principle is however,  not a valid assumption at all scales as from the observations of the large scale structure of universe,
we know that there exists matter inhomogeneities up to the scales of super-clusters of galaxies. Moreover, the success of FLRW cosmology comes with a cost of assuming that
$95\% $ of energy content of our universe is still unknown and is in form of two dark sectors \textit{viz.} ``dark matter'' and ``dark energy''. Existence of dark matter
has been established for a long time. Yet, in spite of numerous ongoing projects there is no direct detection of it till date.
 The  late time cosmic acceleration attributed to dark energy is supported by 
 precise cosmological  observations, such as  Type Ia supernovae~\cite{perl}, cosmic microwave 
 background anisotropies~\cite{sperg}, large scale structure formation~\cite{SDSS},
baryon oscillations~\cite{SDSS2}, and weak lensing~\cite{lens}. The theoretical understanding of 
dark energy  is far from satisfactory, even if one  accepts the 
 cosmological constant $(\Lambda)$~\cite{weinberg}
as an additional constant of nature, with the fine tuning and coincidence problem yet
 to be resolved. 
 
 To address these issues a variety of  approaches that modify or extend the standard picture have been proposed, such as 
scalar field models~\cite{infldarken, scalfield, cope, darkmdarken, joy}, and modified gravity models~\cite{amna1, amna2, amna3, amna4}, to name a few. Most of these approaches are however plagued with problems of their own, primarily to do with
naturalness of the considered scenarios. On the other hand,
from the observations of large scale structure of the universe a question arises that, can the local matter inhomogeneities be linked to  global cosmological
components such as dark matter and dark energy? To address this question one needs to adopt an alternative  approach  to  the
concordance  model  and  consider  the  possibility  that  the dark components  could  emerge  from  a  suitable  averaging  procedure  applicable  to
inhomogeneous  sources. This approach where the effect of deviations from exact homogeneity and isotropy on the average
expansion is studied is called the backreaction approach.   
 Such an approach generalizes the cosmological principle, and considers the universe to be only statistically homogeneous and isotropic at very large scales. 
 
 Several
 averaging techniques~\cite{Zal1:1992Zal2:1993, alan1, Buchert:1999er,Kor:2010,tavakol,Skarke} have been developed to study the relevance
 of backreaction on the average large scale cosmological dynamics and to describe the late time cosmic evolution\cite{alan2, alan3}. 
One of them is Buchert's approach which relies of an average of scalar quantities over spatial
hypersurfaces~~\cite{buchert2, buchert3, weigand, buchert4, rasanen, wiltshire, bose}. Based on the framework developed by Buchert it has been shown~\cite{weigand, buchert4, rasanen}
that backreaction from inhomogeneities could lead to an accelerated expansion of the present universe. Application of the
Buchert framework could however be responsible for an apparent acceleration due to the difference in lapse of time for the underdense and overdense regions~\cite{wiltshire}.
Extrapolation of the Buchert model further predicts the slowing down of the present acceleration at a future era, with possible avoidance of the future big rip~\cite{bose}.
The impact of backreaction from inhomogeneities provides an interesting platform for investigating
the issue of acceleration during the current epoch without invoking modifications
to either GR or the standard model of fundamental forces.
 
In the Buchert's framework the backreaction fluid can be modelled as a scalar field as shown in \cite{morf1}. In this analogy the
effective energy density and an effective pressure corresponding to the scalar field can be obtained from the kinematic backreaction and the averaged curvature term,
which appears as additional terms compared with the standard 
FLRW dynamics.  At scales much larger than the scale of inhomogeneities, the dynamics of the  the global domain is then analogous 
to that of a two-fluid universe, one component being matter, the other one the 
effective scalar field due to backreaction. Scalar field cosmologies~\cite{scalfield} have been
widely studied in the quest of understanding dark energy, since they 
provide the scope of a unified framework for explanation of early inflation with dark 
energy~\cite{infldarken}
on the one hand, and also a possibility of a unified view of the presently
dominant dark sector~\cite{darkmdarken}. In a recent work the backreaction fluid has been constructed considering
a bulk viscous fluid which gives a unified description of dark matter and dark energy~\cite{morf2}.

In this paper we investigate the backreaction fluid to see whether it can act as a source of dark energy thereby giving rise to a model
analogous to scalar field cosmologies in the standard FRW model. We see a number of advantages in such description. First, the effective scalar field has a well justified status in this
theory, and
we do not have to consider it as an additional source arising from an
extraneous field theory. Second, the inhomogeneities encoded in the backreaction term fix the potential, thereby determining the evolution of the scalar field in it,
and therefore, the model does not suffer from phenomenological parametrization. 
Third, the correspondence allows a realistic reinterpretation of phenomenological 
models involving scalar fields, as a large variety of potentials are allowed in such models and there is no way in which one can restrict these potentials. By 
making backreaction responsible for the late time cosmic acceleration the scalar field models can be reinvestigated using a fixed potential. Fourth, the cosmological
constant can arise  as a particular solution of the averaged background without being present in the Einstein's equations.

In the present work we construct a scalar field model
due to backreaction, which is analogous to the quintessence field model of dark 
energy. Our focus here is to study the future evolution of the presently observed
accelerating universe. In order to do so, we first
determine the potential of the effective quintessence field in a two scale model
of inhomogeneities within the Buchert framework. We then study the evolution  of the effective scalar field in the context of the standard FRW background at scales where 
cosmological principle is assumed to hold true. We perform  observational analysis on the model and see how the parameters of the model are constrained.

The plan of the paper is as follows. In the next section we present a brief overview of  Buchert's formalism of backreaction involving partitioning of the global domain into
overdense and underdense regions. In Section III we present our two scale model with
the emergent scalar field. The dynamics of the effective scalar field are presented
in Section IV. In Section V we perform observational analysis of our model by
considering the data from three different experiments, {\it viz.}, Type Ia Supernovae, Baryon Acoustic Oscillations, and Cosmic Microwave Background. We conclude with a 
brief summary of our results in Section VI.

\section{Buchert's backreaction formalism}

Assuming the universe to be filled with an irrotational fluid of dust, the space\textendash{}time
is foliated into flow-orthogonal hypersurfaces with an inhomogeneous $3$-metric  $g_{ij}$ and the line-element
\begin{equation}
ds^{2}=-dt^{2}+g_{ij}dX^{i}dX^{j},
\label{eq:1}
\end{equation}
where  $t$ is the proper time of the hypersurfaces and $X^{i}$ are the spatial coordinates in the hyper-surfaces
of constant $t$. This framework is applicable to perfect fluid matter models. The averaged  equations for volume expansion
and volume acceleration for a compact domain ${\mathcal{D}}$  from  this  approach
are  those  found  by   averaging  the  Hamiltonian,  Raychaudhuri  and
conservation equations\cite{Buchert:1999er,buchert2,buchert3,weigand,buchert4}, and are given by
\begin{equation}
3\frac{\ddot{a}_{\mathcal{D}}}{a_{\mathcal{D}}}= -4\pi G\left\langle \rho\right\rangle _{\mathcal{D}}+\mathcal{Q}_{\mathcal{D}}
+\Lambda,
\label{eq:2}
\end{equation}
\begin{equation}
3H_{\mathcal{D}}^{2} = 8\pi G\left\langle \rho\right\rangle _{\mathcal{D}}-\frac{1}{2}\mathcal{\left\langle R\right\rangle }_{\mathcal{D}}
-\frac{1}{2}\mathcal{Q}_{\mathcal{D}}+\Lambda,
\label{eq:3}
\end{equation}

\begin{equation}
\dot{\left\langle\rho\right\rangle}_{\mathcal{D}} + 3H_{\mathcal{D}} \left\langle \rho\right\rangle_{\mathcal{D}}=0,
\label{eq:integrability}
\end{equation}
where $a_{\mathcal{D}}$ is the volume scale factor of the global domain $\mathcal{D}$, $\left\langle \rho\right\rangle_\mathcal{D}$  is the
average matter density of the domain, $\left\langle \mathcal{R}\right\rangle_\mathcal{D}$ is the is the averaged three curvature $<\ ^3 R>_{\mathcal{D}}$,
and $H_{\mathcal{D}}$ is the  Hubble rate. The angular brackets denote a volume average throughout
the region ${\mathcal{D}}$, and  $\mathcal{Q_{D}}$ is the the  backreaction term that quantifies the averaged effect of inhomogeneities in the background. It is defined
as 
\begin{equation}
\mathcal{Q_{D}}=\frac{2}{3}\left(\left\langle \theta^{2}\right\rangle _{\mathcal{D}}-\left\langle \theta\right\rangle _{\mathcal{D}}^{2}\right)-2\left\langle\sigma^{2}\right\rangle_{\mathcal{D}},
\label{eq:5}
\end{equation}
where $\theta$ is the volume expansion rate and $\sigma^{2}=1/2\sigma_{ij}\sigma^{ij}$
is the squared rate of shear. The volume average of a 
scalar quantity $\left\langle f\right\rangle {}_{\mathcal{D}}(t)$
over the rest mass preserving domain $\mathcal{D}$,  is given by
\begin{equation}
\left\langle f\right\rangle {}_{\mathcal{D}}(t)=\frac{1}{|\mathcal{D}|_{g}}\int_{\mathcal{D}}fd\mu_{g},
 \label{eq:6}
\end{equation}
where $|\mathcal{D}|_{g}=\int_{\mathcal{D}}d\mu_{g}$ is the volume of the domain and $d\mu_{g}=\sqrt{^{(3)}g(t,X^{1},X^{2},X^{3})}dX^{1}dX^{2}dX^{3}$.
The volume scale factor $a_{\mathcal{D}}$ is defined as:
\begin{equation}  
a_{\mathcal{D}}(t)=\left(\frac{|\mathcal{D}|_{g}}{|\mathcal{D}_{0}|_{g}}\right)^{1/3},
\end{equation}
where $|\mathcal{D}_{0}|_{g}$ is the  volume of the domain at a reference  time $t_0$
which we may take as the present time. The quantities
$\mathcal{Q_{\mathcal{D}}}$ and $\left\langle \mathcal{R}\right\rangle_\mathcal{D}$ obey the consistency relation
\begin{equation}
\frac{1}{a_{\mathcal{D}}^{6}}\partial_{t}\left(a_{\mathcal{D}}^{6}\mathcal{Q}_{\mathcal{D}}\right)
+\frac{1}{a_{\mathcal{D}}^{2}}\partial_{t}\left(a_{\mathcal{D}}^{2}\mathcal{\left\langle R\right\rangle }_{\mathcal{D}}\right)=0,
\label{eq:integrability1}
\end{equation}
which
connects \eqref{eq:2} and \eqref{eq:3}, assuming that the expansion equation as integral of the acceleration equation. 
While the mass conservation law for matter is sufficient for the case of a smooth 
universe, the additional
equation \eqref{eq:integrability1}   connects the averaged intrinsic and extrinsic curvature invariants when the effect of inhomogeneities on the 
background metric are taken into consideration.
The new second conservation equation \eqref{eq:integrability1} describes an interaction between structure 
formation and the background curvature. The dynamical coupling of $\mathcal{Q_{D}}$ to 
the average scalar curvature $\left\langle \mathcal{R}\right\rangle_\mathcal{D}$  implies that the temporal behaviour
of the averaged curvature is different from that of the standard FRW constant curvature model where it is
 restricted to an $a_{\mathcal{D}}^{-2}$ behaviour, whereas in the present case
 it is more dynamical since it can be any function of $a_{\mathcal{D}}$.
 
At scales much larger than the scale of inhomogeneites, the global domain is assumed to be homogeneous, and the above equations can formally be recast into standard
Friedmann equations for an effective perfect fluid energy momentum tensor
with new effective sources as~\cite{morf1}:
\begin{eqnarray}
\rho_{eff}^{\mathcal{D}}=\left\langle \rho\right\rangle_{\mathcal{D}}
-\frac{1}{16\pi G}\mathcal{Q_{D}}-\frac{1}{16\pi G}\left\langle \mathcal{R}\right\rangle_\mathcal{D};\nonumber\\
 P_{eff}^{\mathcal{D}}=-\frac{1}{16\pi G}\mathcal{Q_{D}}+\frac{1}{16\pi G}\left\langle \mathcal{R}\right\rangle_\mathcal{D},
 \label{eq:9}
\end{eqnarray}
such that
\begin{equation}
3\frac{\ddot{a}_{\mathcal{D}}}{a_{\mathcal{D}}}= -4\pi G (\rho_{eff}^{\mathcal{D}}+3P_{eff}^{\mathcal{D}})
+\Lambda,
\label{eq:10}
\end{equation}
\begin{equation}
3H_{\mathcal{D}}^{2} = 8\pi G\rho_{eff}^{\mathcal{D}}+\Lambda,
\label{eq:11}
\end{equation}
\begin{equation}
\dot\rho_{eff}^{\mathcal{D}} + 3H_{\mathcal{D}} (\rho_{eff}^{\mathcal{D}}+P_{eff}^{\mathcal{D}})=0\,.
\label{eq:integrability2}
\end{equation}
Eqs.\eqref{eq:10}, \eqref{eq:11} and \eqref{eq:integrability2} correspond
to the eqs.\eqref{eq:2}, \eqref{eq:3} and \eqref{eq:integrability} respectively.

Since the scalar functions are domain dependent, following \cite{weigand}  the homogeneous global domain 
is partitioned into noninteracting
 subregions $\mathcal{F}_{\ell}$ which themselves consist of
elementary space entities $\mathcal{F}_{\ell}^{(\alpha)}$ that is associated with some averaging length scale. Mathematically 
$\mathcal{D}=\cup_{\ell}\mathcal{F}_{\ell}$, where $\mathcal{F}_{\ell}=\cup_{\alpha}\mathcal{F}_{\ell}^{(\alpha)}$
and $\mathcal{F}_{\ell}^{(\alpha)}\cap\mathcal{F}_{m}^{(\beta)}=\emptyset$
for all $\alpha\neq\beta$ and $\ell\neq m$. One can construct the average of the scalar
 function $f$ on the domain $\mathcal{D}$ as in eqn.\eqref{eq:2}  summing all the averages 
 $f$ on the subregions $\mathcal{F}_{\ell}$ as, 
\begin{equation}
\left\langle f\right\rangle _{\mathcal{D}}=\underset{\ell}{\sum}|\mathcal{D}|_{g}^{-1}\underset{\alpha}{\sum}\int_{\mathcal{F}_{\ell}^{(\alpha)}}fd\mu_{g}=\underset{\ell}{\sum}\lambda_{\ell}\left\langle f\right\rangle _{\mathcal{F}_{\ell}},
\end{equation}
where $\lambda_{\ell}=|\mathcal{F}_{\ell}|_{g}/|\mathcal{D}|_{g}$,
is the volume fraction of the subregion $\mathcal{F}_{\ell}$. The scalar quantities $\left\langle \rho\right\rangle_{\mathcal{D}}$, $<\mathcal{R}>_{\mathcal{D}}$ and $H_{\mathcal{D}}$ follow the above equation
except  $\mathcal{Q_{D}}$, since from its definition in eqn.\eqref{eq:3} we notice that
due to the term $\left\langle \theta\right\rangle _{\mathcal{D}}^{2}$ it  does not
split in a simple manner. Instead, the correct formula turns out to be 
\begin{equation}
\mathcal{Q_{D}}=\underset{\mathcal{D}}{\mathcal{\sum}}\lambda_{\ell}\mathcal{Q}_{\ell}+3\underset{\ell\neq m}{\sum}\lambda_{\ell}\lambda_{m}\left(H_{\ell}-H_{m}\right)^{2}\label{eq:4},
\end{equation}
where $\mathcal{Q}_{\ell}$ and $H_{\ell}$ are defined in the subregion $\mathcal{F}_{\ell}$
in the same way as $\mathcal{Q}_{\mathcal{D}}$ and $H_{\mathcal{D}}$
are defined in $\mathcal{D}$. The shear part $\left\langle \sigma^{2}\right\rangle _{\mathcal{F}_{\ell}}$
is completely absorbed in $\mathcal{Q}_{\ell}$ , whereas the variance
of the local expansion rates$\left\langle \theta^{2}\right\rangle _{\mathcal{D}}-\left\langle \theta\right\rangle _{\mathcal{D}}^{2}$
is partly contained in $\mathcal{Q}_{\ell}$ and partly generates the
extra term $3\sum_{\ell\neq m}\lambda_{\ell}\lambda_{m}\left(H_{\ell}-H_{m}\right)^{2}$.
This is because the  variance $\left\langle \theta^{2}\right\rangle _{\mathcal{\mathcal{F}_{\ell}}}-\left\langle \theta\right\rangle _{\mathcal{F}_{\ell}}^{2}$
that is present in the backreaction term $\mathcal{Q}_{\ell}$ of a subregion takes care only
of the points inside the subregion $\mathcal{F}_{\ell}$. To take into account of all the variance
 that comes from combining points of the subregion $\mathcal{F}_{\ell}$
with others in $\mathcal{F}_{m}$, the extra term which contains the averaged
Hubble rate in above equation emerges. It is to be noted here that this
formulation of the backreaction only
holds in the case when there is no overlap between the different subregions.

As the volume of the global domain $\mathcal{D}$ is equal to the sum 
of volumes of all individual sub domains such that $|\mathcal{D}|_{g}=\sum_{\ell}|\mathcal{F}_{\ell}|_{g}$,
the corresponding scale-factor
$a_{\ell}$ for each of the subregions $\mathcal{F}_{\ell}$ can be
defined as
\begin{equation}
a_{\mathcal{D}}^{3}=\sum_{\ell}\lambda_{\ell_{i}}a_{\ell}^{3}\label{globscale},
\end{equation}
where $\lambda_{\ell_{i}}=|\mathcal{F}_{\ell_{i}}|_{g}/|\mathcal{D}_{i}|_{g}$
is the initial volume fraction of the subregion $\mathcal{F}_{\ell}$.
If we now twice differentiate this equation with respect to the foliation
time and use the result for $\dot{a}_{\ell}$ from \eqref{eq:3},
we then get the expression that relates the acceleration of the global
domain to that of the subdomains: 
\begin{equation}
\frac{\ddot{a}_{\mathcal{D}}}{a_{\mathcal{D}}}=\underset{\ell}{\sum}\lambda_{\ell}\frac{\ddot{a}_{\ell}(t)}{a_{\ell}(t)}+
\underset{\ell\neq m}{\sum}\lambda_{\ell}\lambda_{m}\left(H_{\ell}-H_{m}\right)^{2}.\label{eq:5b}
\end{equation}
From the above equation we notice that even if the subdomains decelerate individually,
the second term in the equation may counterbalance the first
one to lead to global accelerated
expansion.

\section{Two scale partitioning of the Universe and the emerging scalar field}

For our purpose we consider the universe to be partitioned into two subdomains: (i) 
all overdense regions are clubbed together into the domain of 
over density called wall $\mathcal{M}$, and (ii) all underdense regions are clubbed
together into the domain of under density
called void $\mathcal{E}$, such that $\mathcal{D}=\mathcal{M}\cup\mathcal{E}$. 
Now, from the Buchert's formalism  as explained in the previous section 
one obtains $H_{\mathcal{D}}=\lambda_{\mathcal{M}}H_{\mathcal{M}}+\lambda_{\mathcal{E}}H_{\mathcal{E}}$,
with similar expressions for $\left\langle \rho\right\rangle _{\mathcal{D}}$
and $\left\langle \mathcal{R}\right\rangle _{\mathcal{D}}$. Simplifying eqn.\eqref{eq:5} for the
two domain case considered here, we get
\begin{equation}
\frac{\ddot{a}_{\mathcal{D}}}{a_{\mathcal{D}}}=\lambda_{\mathcal{M}}\frac{\ddot{a}_{\mathcal{M}}}{a_{\mathcal{M}}}
+\lambda_{\mathcal{E}}\frac{\ddot{a}_{\mathcal{E}}}{a_{\mathcal{E}}}+2\lambda_{\mathcal{M}}\lambda_{\mathcal{E}}(H_{\mathcal{M}}-H_{\mathcal{E}})^{2},
\label{eq:6b}
\end{equation}
where $\lambda_{\mathcal{M}}=|\mathcal{M}|/|\mathcal{D}|$ and $\lambda_{\mathcal{E}}=|\mathcal{E}|/|\mathcal{D}|$ are the 
volume fractions of wall and void respectively. 
Since $\sum_{\ell}\lambda_{\ell}=1$, therefore,
\begin{equation}
\lambda_{\mathcal{M}}+\lambda_{\mathcal{E}}=1.
\label{eq:cons}
\end{equation}

 We assume different behaviours for the scale factors of the two sub regions.
The wall $\mu$ is considered to be a closed region with a positive curvature and a deceleration parameter 
 $q_{\mathcal{M}}=-\ddot{a}_{\mathcal{M}}/a_{\mathcal{M}} H_{\mathcal{M}}^{2} >.5$. The scale factor $a_{\mathcal{M}}$ 
 in terms of the time $t$ parametrized by the angle $\theta$ is given by:
\begin{equation}
a_{\mathcal{M}}=\frac{q_{\mathcal{M}}}{2q_{\mathcal{M}}-1}(1-Cos(\theta)),
\end{equation}
\begin{equation}
t=\frac{q_{\mathcal{M}}}{2q_{\mathcal{M}}-1}(\theta-Sin(\theta)),
\end{equation}
with
$\theta$ varying from $0$ to $2\pi$. The void is considered to be a flat (zero intrinsic curvature domain) with its scale factor $a_{\mathcal{E}}$ assumed  to 
follow a power law:
\begin{equation}
a_{\mathcal{E}}=c_{\mathcal{E}}t^{\beta},
\end{equation}
where $\beta$ and $c_{\mathcal{E}}$ are constants. From eqn.\eqref{eq:6} the acceleration 
equation of the global domain is given by
\begin{equation}
\frac{\ddot{a}_{\mathcal{D}}}{a_{\mathcal{D}}}=\lambda_{\mathcal{M}}(-q_{\mathcal{M}} H_{\mathcal{M}}^2)
+\lambda_{\mathcal{E}}\frac{\beta(\beta-1)}{t^{2}}
  +2\lambda_{\mathcal{M}}\lambda_{\mathcal{E}}\left(H_{\mathcal{M}}-\frac{\beta}{t}\right)^{2}.
  \label{eq:19a}
\end{equation}

Since the volume fractions $\lambda_{\mathcal{M}}$ and $\lambda_{\mathcal{E}}$ 
are constrained by 
eqn.\eqref{eq:cons},  determination of one fixes the other.  The volume
fraction of the void subdomain $\mathcal{E}$ is given by
$\lambda_{\mathcal{E}}=\frac{|\mathcal{E}|}{|\mathcal{D}|}$,
which can be rewritten in terms of the corresponding scale factor
as $\lambda_{\mathcal{E}}=\frac{a_{\mathcal{E}}^{3}|\mathcal{E}|}{a_{\mathcal{D}}^{3}|\mathcal{D}|}$,
which in turn gives us:
\begin{equation}
\lambda_{\mathcal{E}}=k \frac{t^{3\beta}}{a_{\mathcal{D}}^{3}},
\end{equation}
where where $k=\frac{\lambda_{\mathcal{E}_{0}}a_{\mathcal{D}_{0}}^{3}}{t_{0}^{3\beta}}$
is a constant and $\lambda_{\mathcal{E}_{0}}$, $a_{\mathcal{D}_{0}}$, ${t_{0}}$ are the present 
volume fraction of void region, the present global scale factor and the present time, respectively. 
An N-body
simulation of large scale structure formation~\cite{weigand} indicates the value $\lambda_{\mathcal{M}_{0}}=.09$, and hence using
 eqn.\eqref{eq:cons}, we choose $\lambda_{\mathcal{E}_{0}}= 1-\lambda_{\mathcal{M}_{0}}=.91$.
Eqn.\eqref{eq:19a} can thus be written as:
\begin{eqnarray}
\frac{\ddot{a}_{\mathcal{D}}}{a_{\mathcal{D}}}& = &\left(1-\frac{k t^{3\beta}}{a_{\mathcal{D}}^{3}}\right)(-q_{\mathcal{M}} H_{\mathcal{M}}^2)
+\frac{k t^{3\beta}}{a_{\mathcal{D}}^{3}}\frac{\beta(\beta-1)}{t^{2}}\nonumber \\
 &  & +2\frac{k t^{3\beta}}{a_{\mathcal{D}}^{3}}\left(1-\frac{k t^{3\beta}}{a_{\mathcal{D}}^{3}}\right)\left(H_{\mathcal{M}}-\frac{\beta}{t}\right)^{2}\,.\label{eq:19}
\end{eqnarray}

Now, using eq.\eqref{eq:9}, the potential
$V_{\mathcal{D}}(\phi_{\mathcal{D}})$ of the   effective scalar  field $\phi_{\mathcal{D}}$ may be obtained as follows~\cite{morf1}:
\begin{eqnarray}
\rho_{eff}^{\mathcal{D}}=\left\langle \rho\right\rangle_{\mathcal{D}}+\rho_{\phi}^{\mathcal{D}};
 & & P_{eff}^{\mathcal{D}}=P_{\phi}^{\mathcal{D}},
\label{effrho}
\end{eqnarray}
where,
\begin{eqnarray}
\rho_{\phi}^{\mathcal{D}}=\frac{1}{2}\dot\phi_{\mathcal{D}}^2+ V_{\mathcal{D}};
 & & P_{\phi}^{\mathcal{D}}=\frac{1}{2}\dot\phi_{\mathcal{D}}^2- V_{\mathcal{D}}.
 \label{effphi}
\end{eqnarray}
Setting $\Lambda=0$ in the Eqs.\eqref{eq:10} and  Eqs.\eqref{eq:11}, and using
eqs.\eqref{eq:integrability2},\eqref{effrho} and \eqref{effphi}, we get the potential
of the field to be
\begin{equation}
V_{\mathcal{D}}=\frac{1}{8 \pi G}\frac{\ddot{a}_{\mathcal{D}}} {a_{\mathcal{D}}}+2\left(\frac{\dot{a}_{\mathcal{D}}(t)} {a_{\mathcal{D}}(t)}\right)^2-
\frac{3}{16\pi G}\frac{ \Omega_{m0}H_0^2}{ a_{\mathcal{D}}^3},
\end{equation}
where 
$\Omega_{m0}=\rho_{m0}/3 H_0^2 M_p^2$, $M_p=1/\sqrt{8\pi G}.$
Next, let us define the quantities:
\begin{equation}
 H_0t\rightarrow t_n,\,\, \frac{\phi_{\mathcal{D}}}{M_p}\rightarrow \phi,\,\, 
 \frac{V_{\mathcal{D}} }{H_0^2 M_p^2}\rightarrow V.\label{dim}
\end{equation}
In terms of the above dimensional quantities, the potential becomes:
\begin{equation}
V(\phi(t))=\frac{\ddot{a}_{\mathcal{D}}(t)} {a_{\mathcal{D}}(t)}+2\left(\frac{\dot{a}_{\mathcal{D}}(t)} {a_{\mathcal{D}}(t)}\right)^2-
\frac{3 \Omega_{m0}}{2 a_{\mathcal{D}}(t)^3},
\label{eq:29}
\end{equation}
From Eqs.\ref{eq:9} we obtain the following correspondence:
\begin{eqnarray}
-\frac{1}{8\pi G}\mathcal{Q_{D}}=\dot\phi(t)-V(\phi);& &  
- \frac{1}{8\pi G}\left\langle \mathcal{R}\right\rangle_\mathcal{D}=3 V(t),
\end{eqnarray} 
Inserting the above equations into the integrability condition \ref{eq:integrability1} implies that $\phi(t)$, for $\dot\phi(t)\neq 0$, 
obeys the Klein-Gordon equation:
\begin{equation}
\ddot\phi(t)+3H_{\mathcal{D}}\dot\phi(t)
+\frac{dV(\phi)}{d\phi}=0\\,\label{KG}
\end{equation}
where $t$ is the dimensionless time$(t_n)$ defined in \eqref{dim} (we henceforth drop the subscript for simplicity). 

\begin{figure}
\centerline{\includegraphics[width=80mm,scale=0.9]{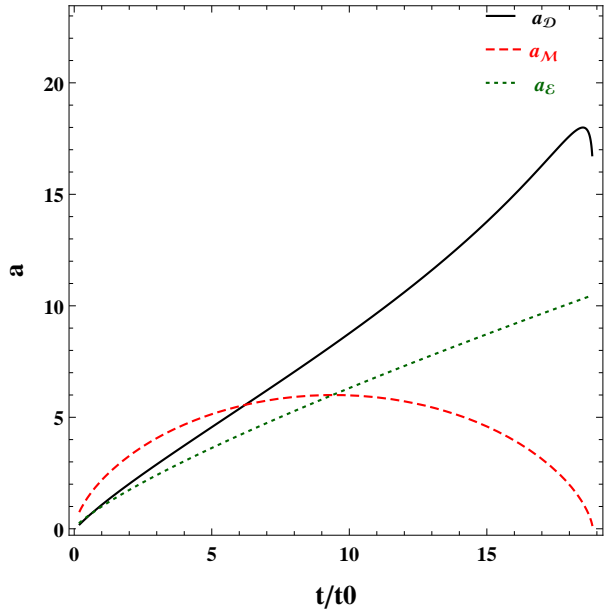} }

\caption{The scale factor of the over-dense subdomain $a_{\mathcal{M}}$(wall), 
the under-dense subdomain $a_{\mathcal{E}}$(void) and 
the  global domain $a_{\mathcal{D}}$ with time(in units of
$t/t_0$ with $t_0$ being the present time) for $q_{\mathcal{M}}=.6$ and $\beta=.7$. The global scale factor $a_{\mathcal{D}}$ first increases
giving a period of acceleration due to the addition of two scale factor $(a_{\mathcal{M}} , a_{\mathcal{E}})$, and then the collapse in $\mathcal{M}$-region
forces the decline of $a_{\mathcal{D}}$ at late times.}

\label{fig1} 
\end{figure}

One may now study the consequence of backreaction due to inhomogeneities in terms of
the formally equivalent 
dynamics of a homogeneous, minimally coupled scalar field. Given this
correspondence, we can
reconstruct the potential in which the field evolves by solving the  
two equations \eqref{eq:19} and \eqref{KG} numerically given the initial conditions.
For this we chose the initial conditions such that at $t/t_0\rightarrow1$ the present
global scale factor $a_{\mathcal{D}_0}=1$. Fig.\ref{fig1} shows the evolution of scale factors 
of the over-dense subdomain $a_{\mathcal{M}}$ (wall), the under-dense subdomain
$a_{\mathcal{E}}$ (void) and the 
 global domain $a_{\mathcal{D}}$ respectively, with time. We notice that the scale factor 
of global domain increases for certain period of time and then decrease in the late future. 

\begin{figure}\centering
\begin{center}
 $\begin{array}{c@{\hspace{.02in}}c}
\multicolumn{1}{l}{\mbox{}}\\ [-0.5cm] &
        \multicolumn{1}{l}{\mbox{}} \\ [-0.5cm]
        \hspace*{-.6in}
        \includegraphics[width=80mm,scale=0.9]{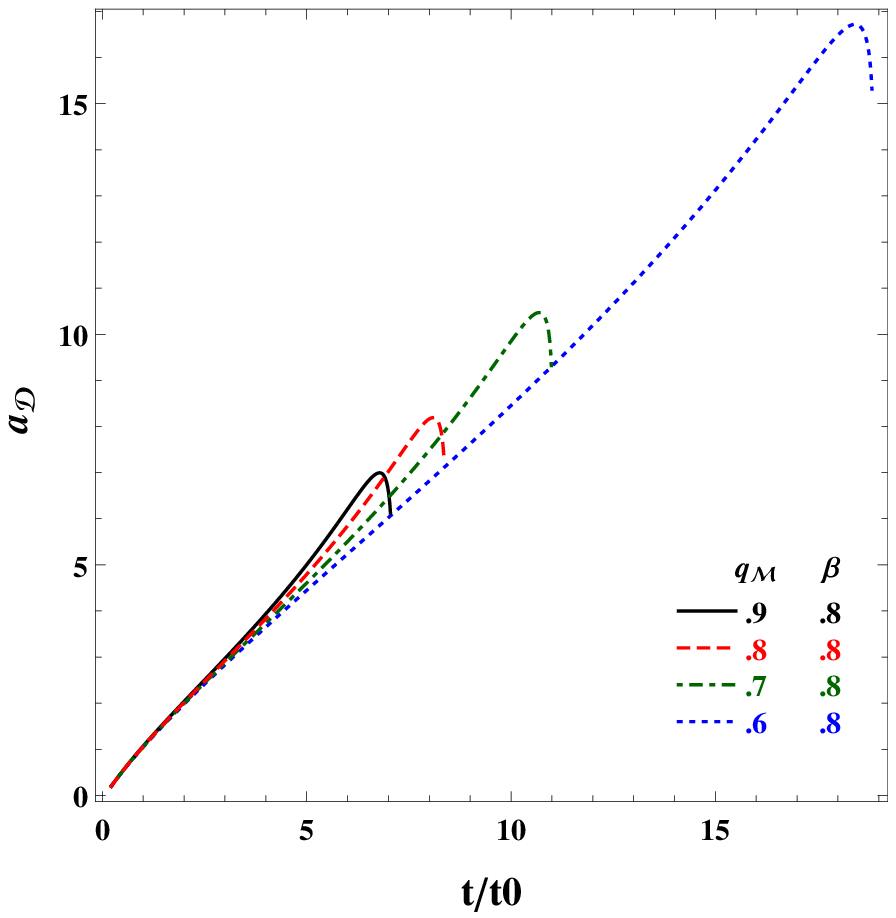} &
                \includegraphics[width=80mm,scale=0.9]{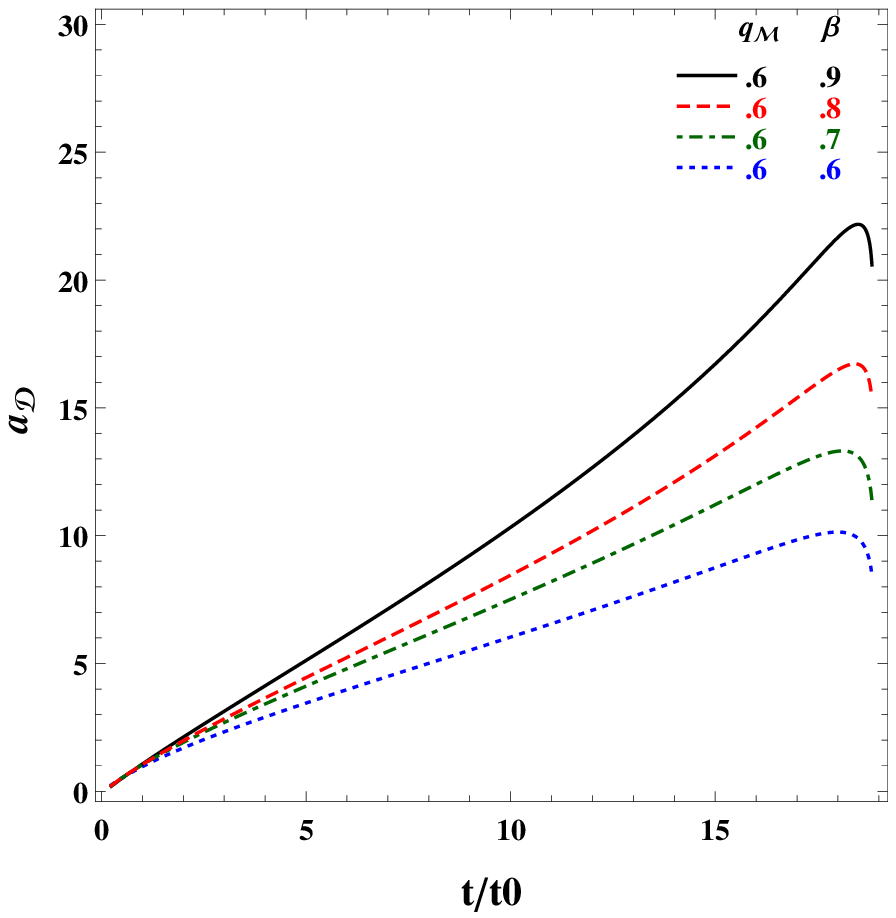} \\ [0.10cm]
\mbox{\bf (a)} & \mbox{\bf (b)}
\end{array}$
\end{center}

\caption{\small The left panel (a): the evolution of the
scale factor of the global domain $a_{\mathcal{D}}$ with time is shown for different values of $q_{\mathcal{M}}$ keeping $\beta$ fixed.
The right panel (b): the evolution of the
scale factor of the global domain $a_{\mathcal{D}}$ with time is shown for different values of $\beta$ keeping $q_{\mathcal{M}}$ fixed. }
\label{fig2}
\end{figure}

We start with a two scale universe where the initial volume fraction of the wall and the void  are equal. With time the void expands whereas the wall contracts. Note that even if finally 
the wall occupies a tiny volume of the global volume, its contraction impacts the evolution
of the global scale factor as a result of which the global Hubble parameter changes its sign. The global Hubble parameter is given
by $H_{\mathcal{D}}=\lambda_{\mathcal{M}}H_{\mathcal{M}}+\lambda_{\mathcal{E}}H_{\mathcal{E}}$.
As time increases, $H_{\mathcal{E}}$ (as per our chosen ansatz given by Eq.(3.5) falls
off rapidly compared to $H_{\mathcal{M}}$ (defined by Eqs.(3.3) and (3.4)).
Thus, the contribution of the decreasing $\lambda_{\mathcal{M}}$ is 
compensated by the comparative behaviours of $H_{\mathcal{M}}$ and $H_{\mathcal{E}}$ to
lead to a net change of sign in the global Hubble parameter at late times.
We would however like to add here that the framework adopted for
our study may not be adequate for investigating the collapsing 
phase of the global domain in the far future. Our assumption of the disjoint nature of the wall and void is likely to be invalidated 
during the collapsing phase in future, as during this the contracting
global volume may entail the wall region overlapping with the void.   From Fig.\ref{fig2} it is evident that time which marks the decrease of $a_{\mathcal{D}}$ depends on the value of 
$q_{\mathcal{M}}$, as it determines the time up to which $a_{\mathcal{M}}$ expands. $\beta$ on the 
other hand determines how much $a_{\mathcal{D}}$ increases before it starts decreasing.
We see that $a_{\mathcal{D}}$ increases with increasing $\beta$.

\begin{figure}[H]
\begin{center}
 $\begin{array}{c@{\hspace{.02cm}}c}
\multicolumn{1}{l}{\mbox{}}\\ [-0.5cm] &
        \multicolumn{1}{l}{\mbox{}} \\ [-0.1cm]
        \hspace*{-.6in}
        \includegraphics[width=80mm,scale=0.9]{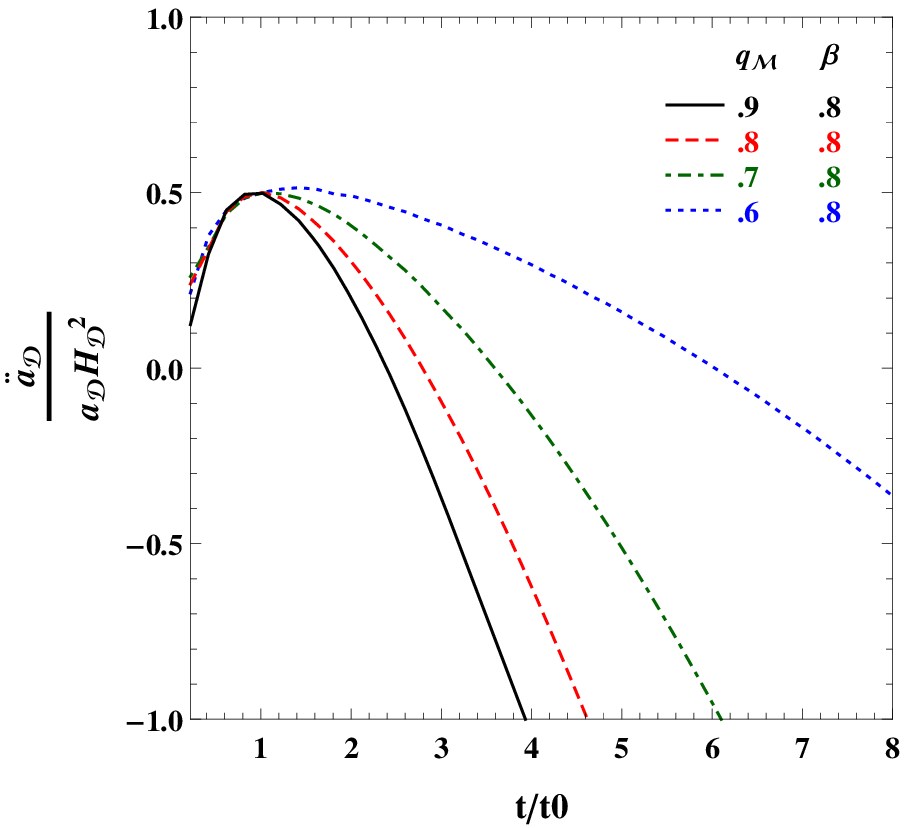} &
                \includegraphics[width=80mm,scale=0.9]{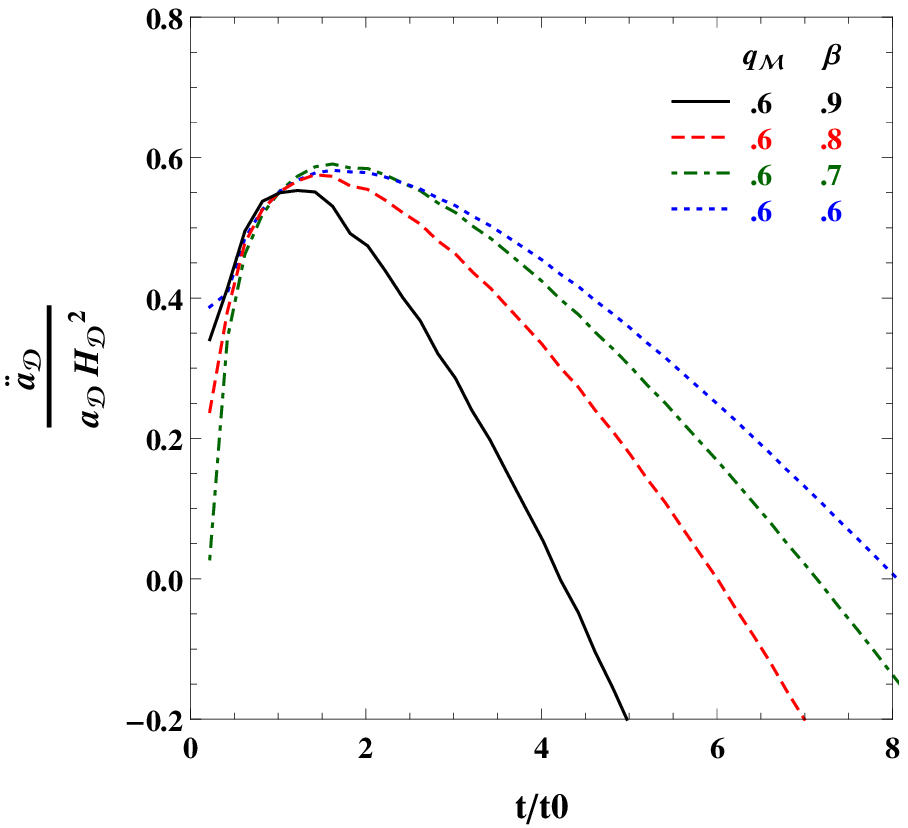} \\ [0.10cm]
\mbox{\bf (a)} & \mbox{\bf (b)}
\end{array}$
\end{center}
\caption{\small The left panel (a): the  dimensionless  global
acceleration  parameter  $\frac{\ddot{a}_{\mathcal{D}}}{a_{\mathcal{D}}H_{\mathcal{D}}^2}$
versus time $(t/t_0)$ is shown for different values of $q_{\mathcal{M}}$ keeping $\beta$ fixed. The right panel (b): the  dimensionless  global
acceleration  parameter  $\frac{\ddot{a}_{\mathcal{D}}}{a_{\mathcal{D}}H_{\mathcal{D}}^2}$
vs. time$(t/t_0)$ is shown for different values of $\beta$ keeping $q_{\mathcal{M}}$ fixed.}
\label{fig3}
\end{figure}

\begin{figure}[H]\centering
\begin{center}
 $\begin{array}{c@{\hspace{.02in}}c}
\multicolumn{1}{l}{\mbox{}}\\ [-0.5cm] &
        \multicolumn{1}{l}{\mbox{}} \\ [-0.5cm]
        \hspace*{-.6in}
        \includegraphics[width=80mm,scale=0.9]{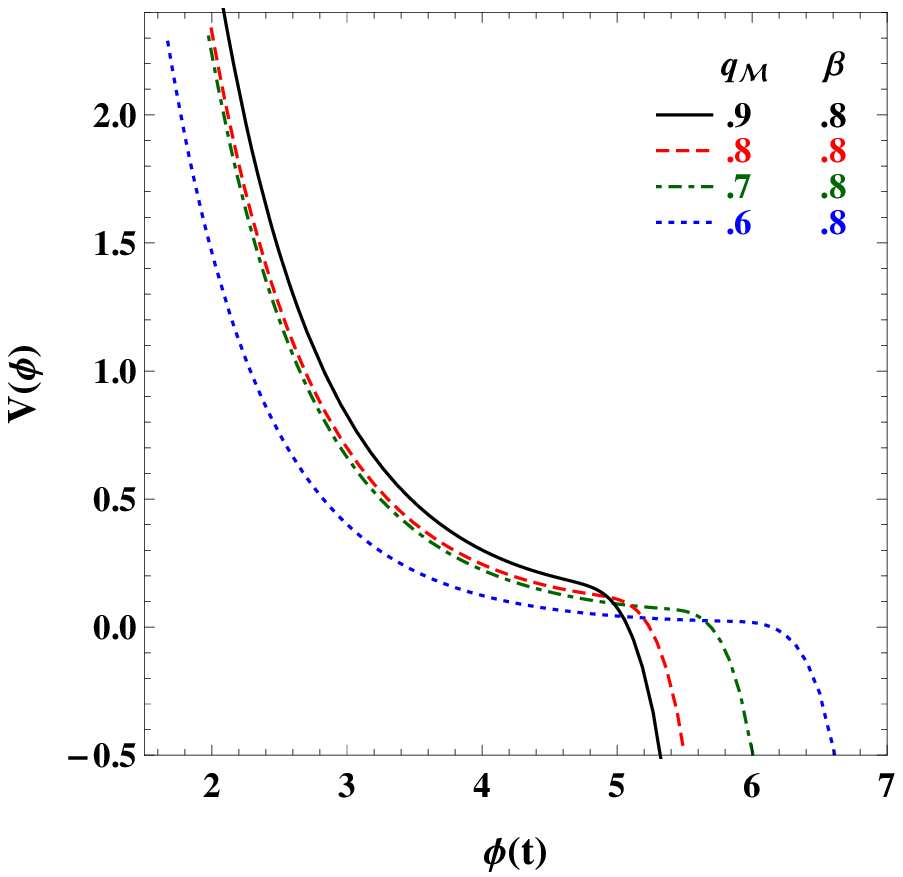} &
                \includegraphics[width=80mm,scale=0.9]{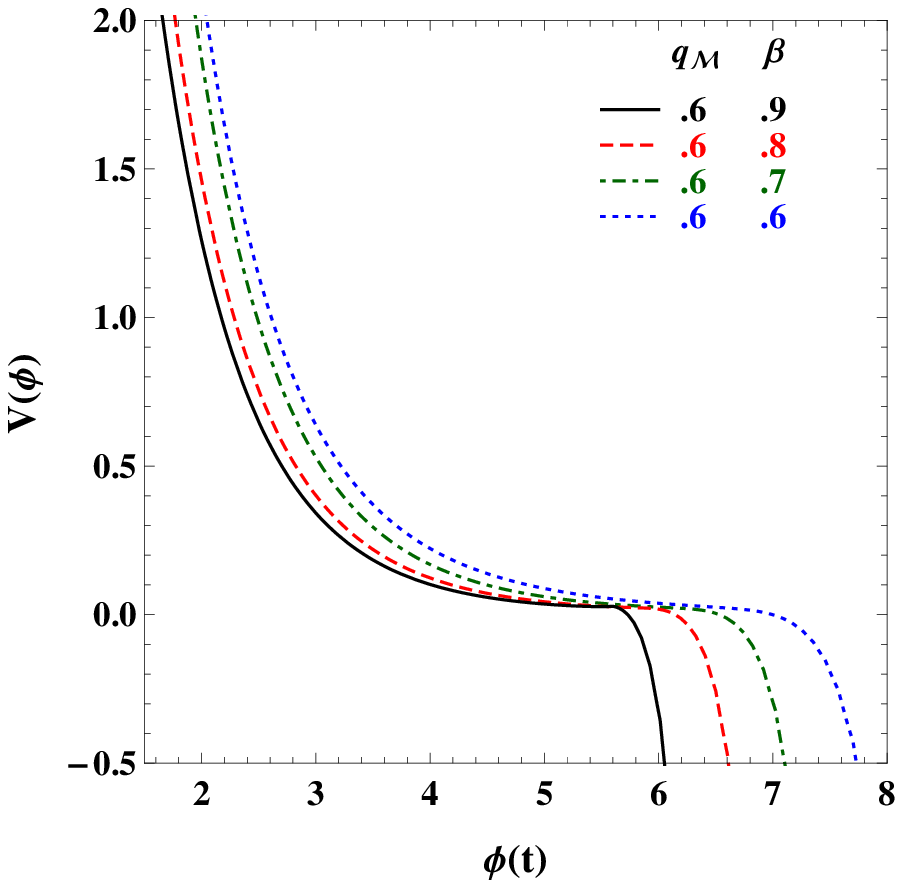} \\ [0.02mm]
\mbox{\bf (a)} & \mbox{\bf (b)}
\end{array}$
\end{center}
\caption{\small The left panel (a): potential $V(\phi)$ versus $\phi(t)$ is shown for different values of $q_{\mathcal{M}}$ keeping $\beta$ fixed.
The right panel (b): potential $V(\phi)$ versus $\phi(t)$  is shown for different values of $\beta$ keeping $q_{\mathcal{M}}$ fixed. }
\label{fig4}
\end{figure}

It is interesting to see what happens to the evolution of the universe
once the present stage of acceleration sets in. 
 For this we scale the present value of acceleration parameter $\frac{\ddot{a}_{\mathcal{D}}}{a_{\mathcal{D}}H_{\mathcal{D}}^2}$ $(=-q_{\mathcal{D}})$to the 
 observed value ($0.5$) and determine the future evolution of the universe. The global acceleration
 for fixed $q_{\mathcal{M}}$ and fixed $\beta$ has been plotted in Fig.\ref{fig3}. We see that the 
 acceleration reaches maximum at present and then it gradually decreases until 
 it becomes negative. This predicts that the present accelerating phase is succeeded
   by a
 future decelerating phase. Next, we study the nature of the potential 
 of the emerging scalar field due to the effect of backreaction.
 In Fig.\ref{fig4} we plot the dimensionless potential $V$ as a function of field $\phi$.
 The nature of potential is unique for the parameter $(q_{\mathcal{M}},\beta)$, and it is 
 free from other parametrization. Further, in the next section we investigate the 
 dynamics of the emerging scalar field in this potential.
 
 \section{Evolution of scalar field in global domain}
 
Assuming that our global domain is large enough such that it can fit the 
 Friedmannian cosmology,
 the effective scalar field formulation of the backreaction problem opens up a
promising avenue for observational analysis of the effect of backreaction. 
We  now exploit the
above correspondence to explicitly reconstruct the scalar field dynamics in a Friedmann universe.
For this we consider the
quintessence scalar field obeying the equation of motion (\ref{eq:29}, \ref{KG}) which mimics the repulsive component
in the cosmological evolution. It is noteworthy that the cosmological 
viability of this scalar field essentially depends on the backreaction parameters 
in the two domains.

In the flat FRW background, the Hubble parameter can be written as:
\begin{equation}
 H^2= \frac{1}{3 M_p^2}(\rho_m +\rho_\phi)\label{h1},
\end{equation}
where $\rho_m$ and $\rho_\phi$ are the energy density of matter and the scalar
field, respectively.
To investigate the dynamics described by the equation of motion of field \eqref{KG}
and the above eqn.\eqref{h1}, it would be convenient to cast them as a
system of first order equations given by
\begin{equation}
 \frac{d\phi}{dN}=\frac{\dot\phi}{h(\phi)},\label{auto1}
\end{equation}
\begin{equation}
 \frac{d\dot\phi}{dN}=-3\dot\phi-\frac{1}{h(\phi)}\frac{dV(\phi)}{d\phi},\label{auto2}
\end{equation}
where,
$\phi$, $\dot\phi$ and $V$ are dimensionless quantities
as defined in Eqn.\eqref{dim} and $N\equiv ln a$. The 
function $h$ is given by
\begin{equation}
 h(\phi)=\sqrt{\Omega_{m0} e^{-3a}+\frac{\dot\phi^2}{6}+\frac{V(\phi)}{3}} \,.
\end{equation}

\begin{figure}[H]
\begin{center}
 $\begin{array}{c@{\hspace{.02cm}}c}
\multicolumn{1}{l}{\mbox{}}\\ [-0.5cm] &
        \multicolumn{1}{l}{\mbox{}} \\ [-0.1cm]
        \hspace*{-.6in}
        \includegraphics[width=80mm,scale=0.9]{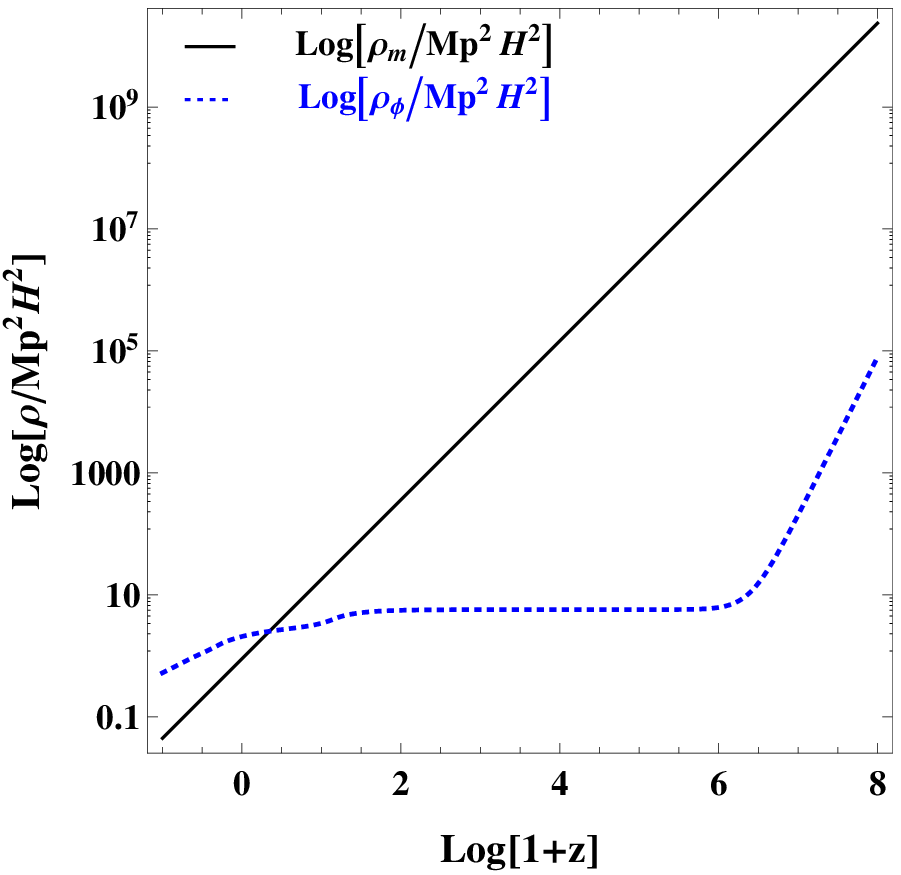} &
                \includegraphics[width=78mm,scale=0.9]{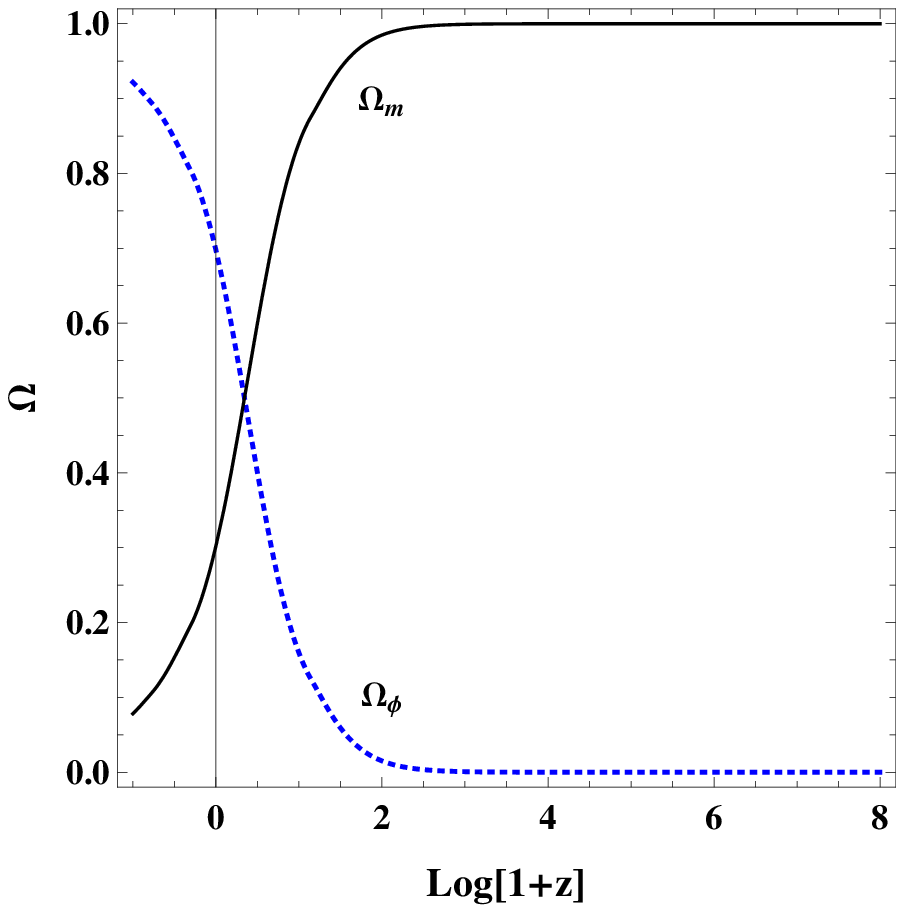} \\ [0.10cm]
\mbox{\bf (a)} & \mbox{\bf (b)}
\end{array}$
\end{center}
\caption{\small The left panel (a): Density 
$\log\Bigl(\frac{\rho}{M_{\rm pl}^2H^2}\Bigr)$ 
for the matter and scalar field are shown for the $q_{\mathcal{M}}=.6$ and $\beta=.9$. The 
right panel (b): density parameters of matter($\Omega_m $),and 
field($\Omega_{\phi}$) are shown for the $q_{\mathcal{M}}=.6$ and $\beta=.9$ }
\label{density}
\end{figure}

\begin{figure}[H]\centering
\begin{center}
 $\begin{array}{c@{\hspace{.02in}}c}
\multicolumn{1}{l}{\mbox{}}\\ [-0.5cm] &
        \multicolumn{1}{l}{\mbox{}} \\ [-0.5cm]
        \hspace*{-.6in}
        \includegraphics[width=78mm,scale=1]{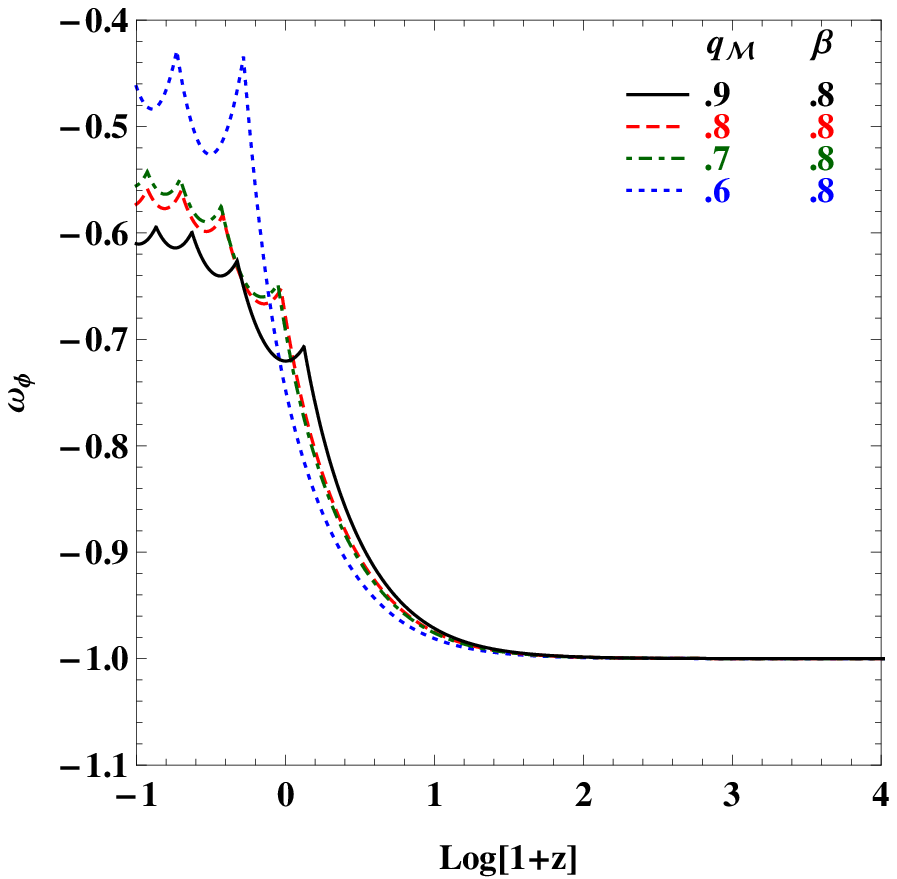} &
                \includegraphics[width=78mm,scale=1]{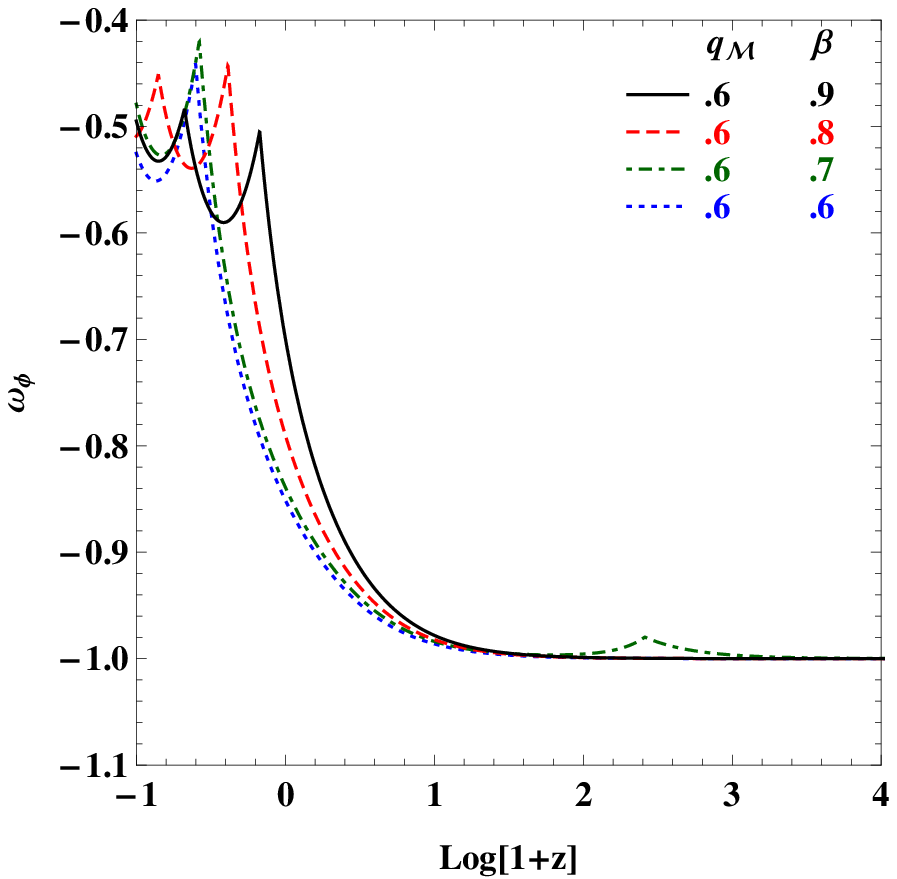} \\ [0.02mm]
\mbox{\bf (a)} & \mbox{\bf (b)}
\end{array}$
\end{center}
\caption{\small The left panel (a): the equation of state of the field $\omega_\phi$ 
for various $q_{\mathcal{M}}$ is shown for fixed $\beta$. The right panel (b): $\omega_\phi$ 
for various $\beta$ is shown for fixed $q_{\mathcal{M}}$.}
\label{eos}
\end{figure}

The initial values for $\phi$($\phi_i$) and
$\dot\phi$  are chosen so as to ensure a viable cosmological model. Since
$\phi_i$ and $\Omega_{m0}$
 are related by the flatness condition $(k = 0, \Omega_{m0} +\Omega_{\phi 0} = 1)$,
an initial value of $\phi_i$ can be fixed by it.  $\phi_i$ controls the deviation 
from the $\Lambda$CDM model. 
In fig.\ref{density} the energy density $(\rho_m,\rho_\phi)$ and the density parameter $(\Omega_m,\Omega_\phi)$ of matter and the
field $\phi$ are plotted. It is seen that during most of the past evolution the energy
density of the scalar field
is sub-dominant, remaining nearly constant for a long time before becoming
 dominant  around the present time when the energy
density of matter falls and becomes comparable to it. 
We plot equation of state parameter of the scalar field $\omega_\phi$ in Fig.\ref{eos} for different
values of $q_\mu$ and $\beta$. We notice that in the past the field is frozen
at $\omega_\phi=-1$, and  begins to evolve only around the present time it. It therefore 
necessarily belongs to the thawing class of scalar field models~\cite{thaw1, thaw2}, where
at early times, the field gets locked at $\omega_\phi=-1$ due to
large Hubble damping and waits for the matter energy
density to become comparable to the field energy density,
which  happens at late times. The field then begins
to evolve toward larger values of $\omega_\phi$ starting from
$\omega_\phi=-1$. From Fig.\ref{eos}  we also see that  the future acceleration   oscillates, and then slows down.
 This decrease in acceleration essentially depends 
on the parameter $q_{\mathcal{M}}$. It may be noted that the oscillation in $\omega_{\phi}$
is a feature of the autonomous Eqs.\eqref{auto1} and \eqref{auto2} considered for the analysis~\cite{sami}.

\section{Data Analysis}
It has been argued that confronting the backreaction models using the cosmological observational data may require more sophisticated approaches~\cite{wilt1,wilt2,wilt3,wilt4} than what one generally does
in case of dark energy models. However, in the present case we have mapped the scalar field originating due to differential scale factors of the local domains
into a large global domain  such that it can fit the Friedmannian cosmology at the relevant
large scales. Therefore, here it is appropriate to use the observational tests where the 
standard method of reduction of the data
assumes the FLRW cosmology.

The evolution of the universe in this set up is controlled by two parameters: one is $\Omega_{m0}$
the present matter energy density of the universe and
the other one is $\phi_{i}$, the initial value of the scalar field.
 But these two parameters
are not completely independent as they can be restrained by
imposing the flatness condition on the space part of the
universe. In this section, we constrain these two parameters with the assumption
of a flat universe by using three types of  observational data, {\it viz.} the Type 1a
Supernovae Union 2.1 compilation~\cite{SnIa}, the Baryon Acoustic Oscillation (BAO) measurement
from the SDSS~\cite{bao1,bao2,bao3}, and
the Cosmic Microwave Background (CMB) measurement given
by WMAP~\cite{cmb}.

The Supernovae type Ia (Sn1a) data, UNION 2.1 compilation 
contains 580 data points, in the form of distance modulus $\mu$ of 
various Sn1a with the redshift $z$.
The distance modulus $\mu$ is defined as
\begin{align}
 \mu=m - M=5 \log D_L+\mu_0\,,
\end{align}
 where $m$ and $M$ are the apparent and absolute magnitudes of the Supernovae respectively. $D_L$  is the luminosity distance defined as
\begin{align}
D_L(z)=(1+z) \int_0^z\frac{H_0dz'}{H(z')}\,,
\end{align}
 and $\mu_0=5 \log\left(\frac{H_0^{-1}}{M_{pc}}\right)+2
5$ is a nuisance parameter which should be marginalized. The corresponding $\chi^2$ is defined as
\begin{align}
\chi_{SN}^2(\mu_0,\theta)=\sum_{i=1}^{580} \frac{\left(\mu_{th}(z_i,\mu_0,\theta)-\mu_{obs}(z_i)\right)^2}{\sigma_\mu(z_i)^2}\,,
\end{align}
where $\mu_{obs} $ is the observational distance modulus,
$\mu_{th}$ is the theoretical distance modulus for the model, 
and $\sigma_{\mu}$ is the error in the distance modulus. $\theta$ represents the parameters
of the model to be constrained and $\mu_0$ is the nuisance parameter.
To get rid of $\mu_0$, we marginalise it as described  
 in Ref.\cite{Lazkoz:2005sp} by defining
 \begin{align}
&A(\theta) =\sum_{i=1}^{580} \frac{\left(\mu_{th}(z_i,\mu_0,\theta)-\mu_{obs}(z_i)\right)^2}{\sigma_\mu(z_i)^2},\\
&B(\theta) =\sum_{i=1}^{580} \frac{\mu_{th}(z_i,\mu_0,\theta)-\mu_{obs}(z_i)}{\sigma_\mu(z_i)^2},\\
&C(\theta) =\sum_{i=1}^{580} \frac{1}{\sigma_\mu(z_i)^2}.
\end{align}
From the above equations one obtains
\begin{align}
\chi_{SN}^2(\theta)=A-\frac{B^2}{C}\,.
\end{align}

We use the large scale structure data~\cite{bao2} for the
BAO co-moving angular-diameter distance ratio $\frac{d_A(z_\star)}{D_V(Z_{BAO})}$,
where $z_\star( \approx 1091)$ is the decoupling time,
$d_A$ is given by $d_A(z)=\int_0^z \frac{dz'}{H(z')}$, 
and $D_V(z)=\left(d_A(z)^2\frac{z}{H(z)}\right)^{\frac{1}{3}}$. Data required for
this analysis is compiled in the table $\ref{baodata}$. 
We calculate  $\chi_{BAO}^2$ as described in Ref.~\cite{bao3}, by defining
\begin{equation}
 \chi_{BAO}^2(\theta)=X_{BAO}^T(\theta) C_{BAO}^{-1} X_{BAO}(\theta).
\end{equation}
where,
\begin{equation}
X_{BAO}(\theta)=\left( \begin{array}{c}
        \frac{d_A(z_\star)}{D_V(0.106)} - 30.95 \\
        \frac{d_A(z_\star)}{D_V(0.2)} - 17.55 \\
        \frac{d_A(z_\star)}{D_V(0.35)} - 10.11 \\
        \frac{d_A(z_\star)}{D_V(0.44)} - 8.44 \\
        \frac{d_A(z_\star)}{D_V(0.6)} - 6.69 \\
        \frac{d_A(z_\star)}{D_V(0.73)} - 5.45
        \end{array} \right),
\end{equation}
and the inverse covariance matrix given by

\begin{align}
C^{-1}=\left(
\begin{array}{cccccc}
 0.48435 & -0.101383 & -0.164945 & -0.0305703 & -0.097874 & -0.106738 \\
 -0.101383 & 3.2882 & -2.45497 & -0.0787898 & -0.252254 & -0.2751 \\
 -0.164945 & -2.45499 & 9.55916 & -0.128187 & -0.410404 & -0.447574 \\
 -0.0305703 & -0.0787898 & -0.128187 & 2.78728 & -2.75632 & 1.16437 \\
 -0.097874 & -0.252254 & -0.410404 & -2.75632 & 14.9245 & -7.32441 \\
 -0.106738 & -0.2751 & -0.447574 & 1.16437 & -7.32441 & 14.5022
\end{array}
\right).
\end{align}

\begin{figure*}\centering
\begin{center}
 $\begin{array}{c@{\hspace{.02in}}c}
\multicolumn{1}{l}{\mbox{}}\\ [-0.5cm] &
        \multicolumn{1}{l}{\mbox{}} \\ [-0.5cm]
        \multicolumn{1}{l}{\mbox{}}\\ [-0.5cm] &
        \multicolumn{1}{l}{\mbox{}} \\ [-0.5cm]
        \hspace*{-.6in}
        \includegraphics[width=80mm,scale=0.9]{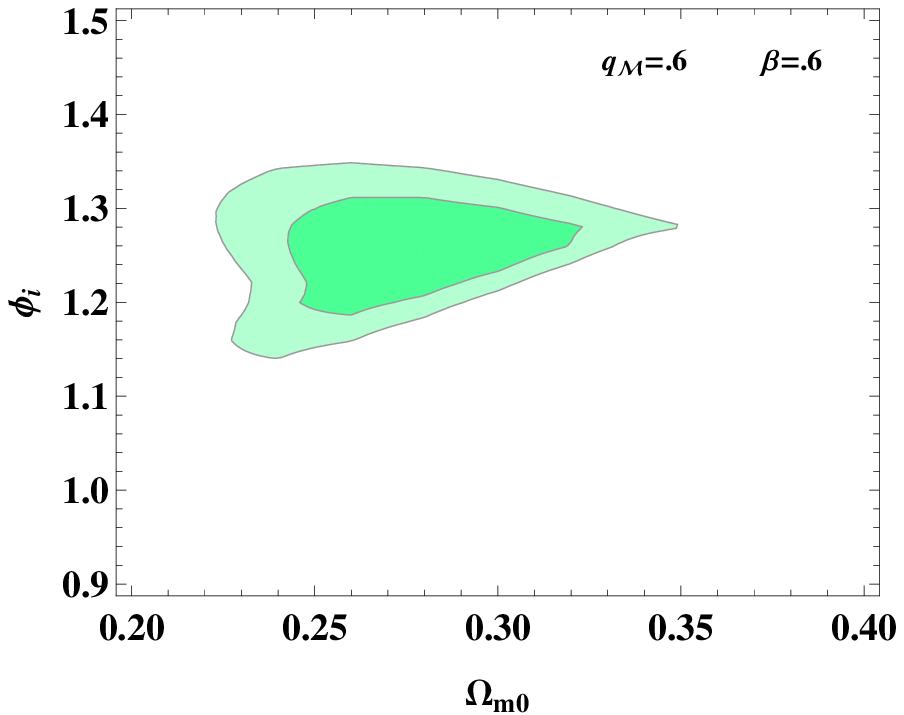} &
                \includegraphics[width=80mm,scale=0.9]{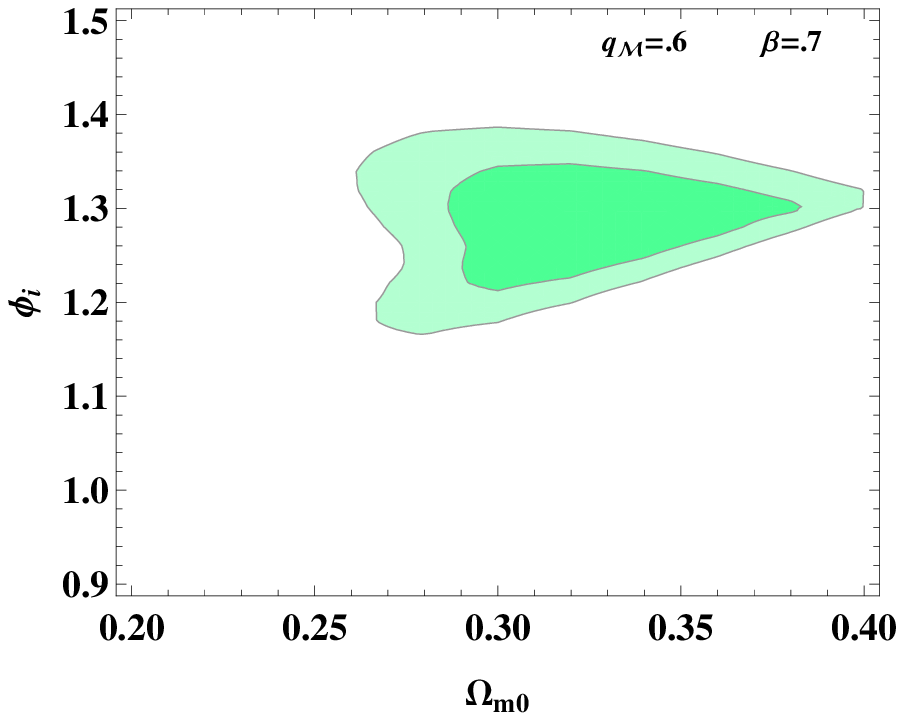} \\ [0.20cm]
      \hspace{-.6in}          \includegraphics[width=80mm,scale=0.9]{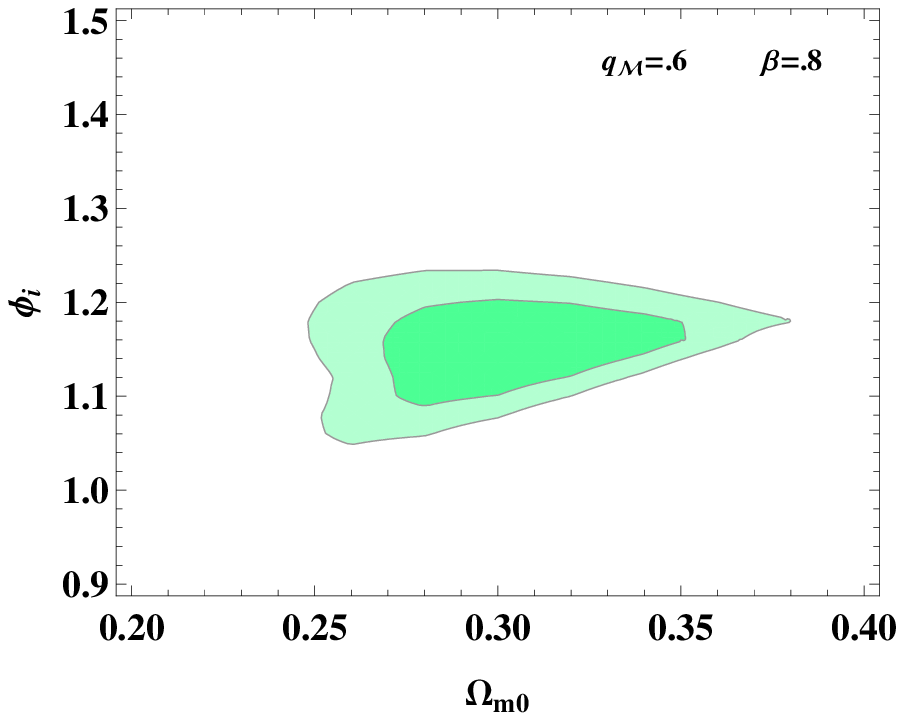} &
                \includegraphics[width=80mm,scale=0.9]{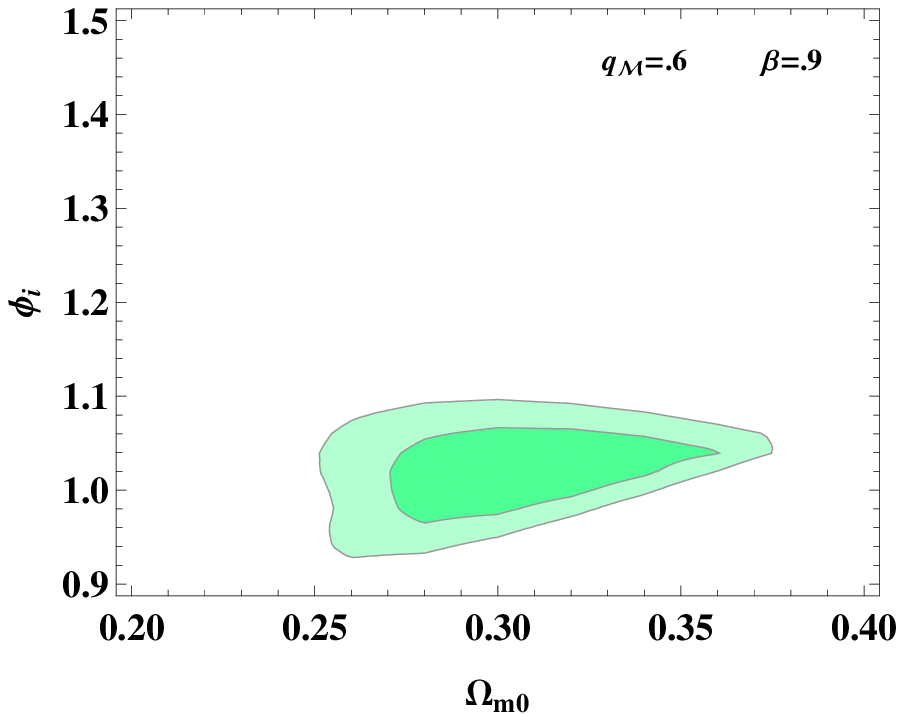} \\ [0.20cm]
\end{array}$
\caption{The 1$\sigma$(dark) and 2$\sigma$(light) likelihood
contours in the $(\Omega_{m0} , \phi_i)$ phase plane for the total
$\chi_{SN+BAO+CMB}^2$ for different values of $\beta$. }
\label{chifixq}
\end{center}
\end{figure*}

\begin{figure*}\centering
\begin{center}
  $\begin{array}{c@{\hspace{.02in}}c}
\multicolumn{1}{l}{\mbox{}}\\ [-0.5cm] &
        \multicolumn{1}{l}{\mbox{}} \\ [-0.5cm]
        \multicolumn{1}{l}{\mbox{}}\\ [-0.5cm] &
        \multicolumn{1}{l}{\mbox{}} \\ [-0.5cm]
        \hspace*{-.6in}
        \includegraphics[width=80mm,scale=0.9]{cont3.eps} &
                \includegraphics[width=80mm,scale=0.9]{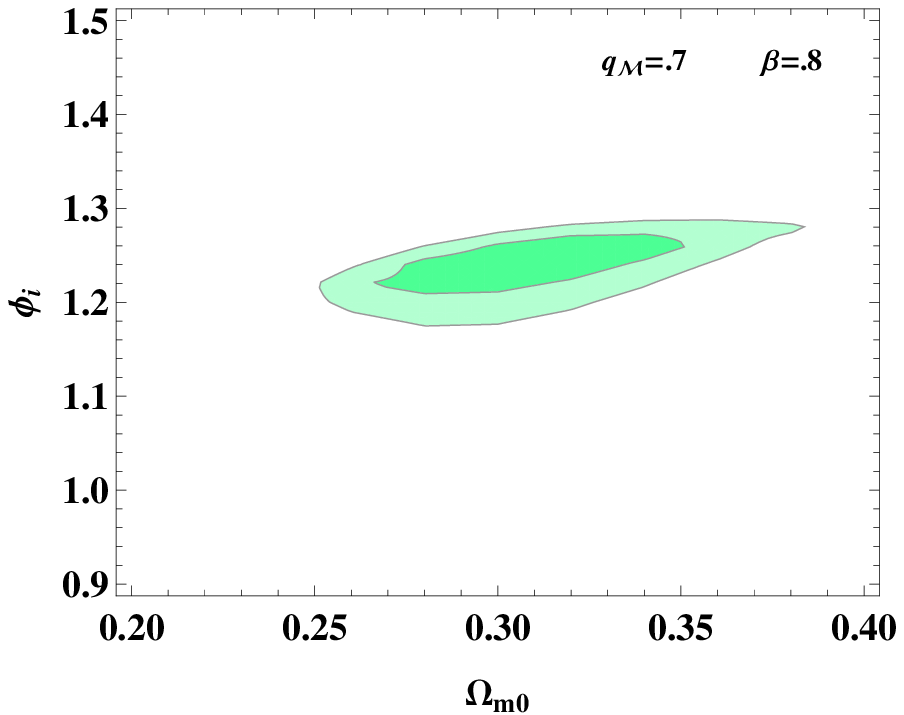} \\ [0.20cm]
      \hspace{-.6in}          \includegraphics[width=80mm,scale=0.9]{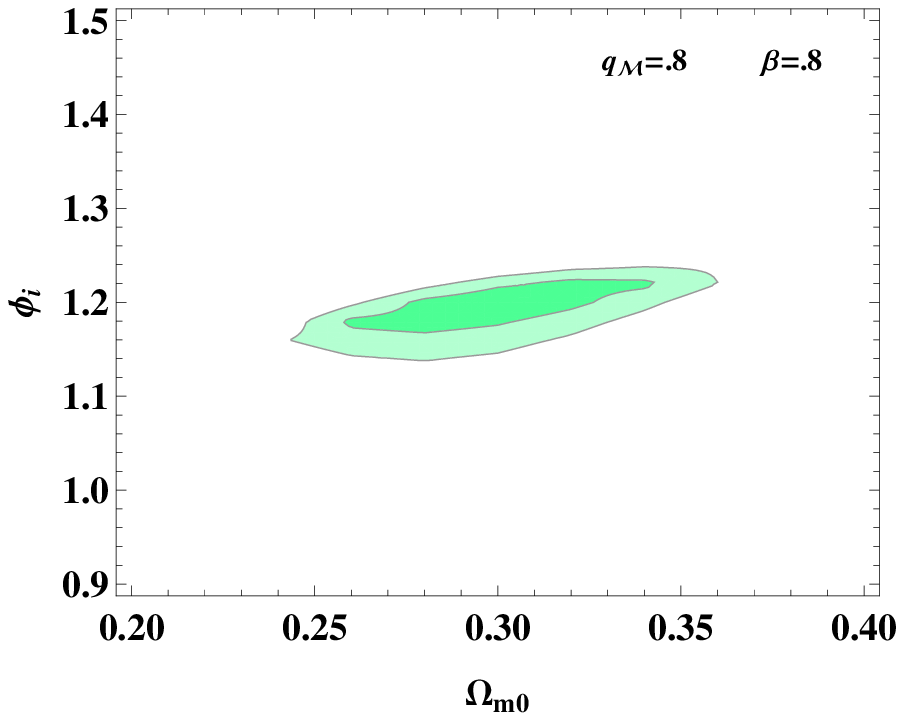} &
                \includegraphics[width=80mm,scale=0.9]{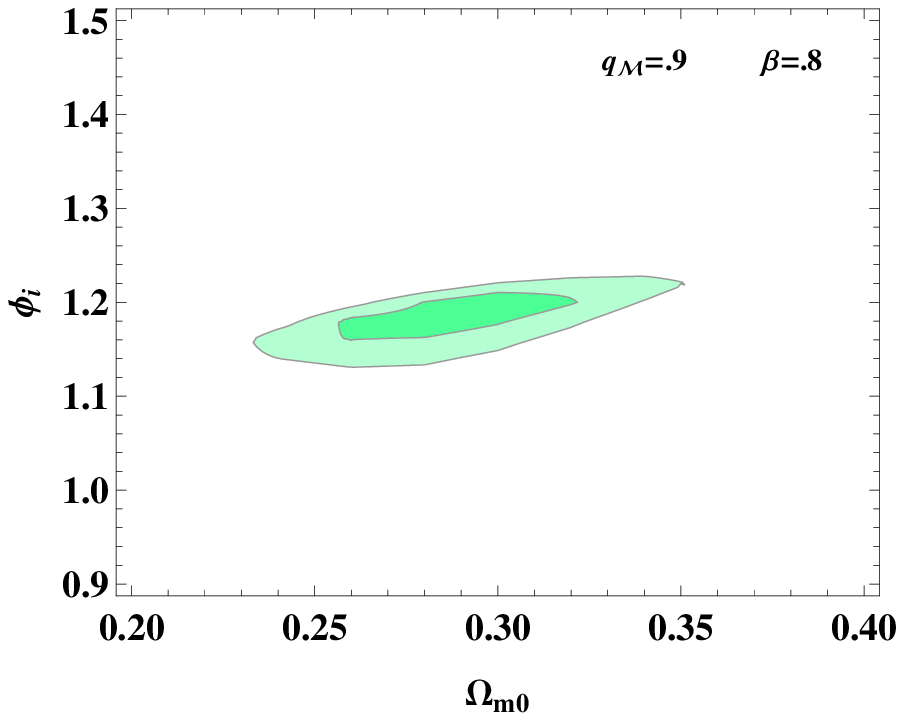} \\ [0.20cm]
\end{array}$
\end{center}
\caption{The 1$\sigma$(dark) and 2$\sigma$(light) likelihood
contours in the $(\Omega_{m0} , \phi_i)$ phase plane for the total
$\chi_{SN+BAO+CMB}^2$ for different values of $q_{\mathcal{M}}$. }
\label{chifixbeta}
\end{figure*}
\begin{center}
\begin{table*}
\begin{tabular}{|c||c|c|c|c|c|c|}
\hline
 $z_{BAO}$  & 0.106  & 0.2 & 0.35 & 0.44 & 0.6 & 0.73\\
\hline \hline
 $\frac{d_A(z_\star)}{D_V(Z_{BAO})}$ &  $30.95 \pm 1.46$ & $17.55 \pm 0.60$ & $10.11 \pm 0.37$ & $8.44 \pm 0.67$ & $6.69 \pm 0.33$ & $5.45 \pm 0.31$  \\
\hline
\end{tabular}
\caption{Values of $\frac{d_A(z_\star)}{D_V(Z_{BAO})}$ for different values of $z_{BAO}$.}
\label{baodata}
\end{table*}
\end{center}

Lastly, for the constraints from CMB we use the CMB shift parameter
\begin{equation}
R=H_0 \sqrt{\Omega_{m0}} \int_0^{1089}\frac{dz'}{H(z')}.
\end{equation}
The $\chi_{CMB}^2$ is defined as
\begin{align}
 \chi_{CMB}^2(\theta)=\frac{(R(\theta)-R_0)^2}{\sigma^2}\,,
\end{align}
where, $R_0=1.725 \pm 0.018$~\cite{cmb}.

The total $\chi_{total}^2$ combining the three data sets is given by
\begin{equation}
 \chi_{total}^2(\theta)=\chi_{SN}^2(\theta)+\chi_{BAO}^2(\theta)+\chi_{CMB}^2(\theta)\,.
\end{equation}
We  carry out this analysis on our model parameters namely, $\phi_i$ and $\Omega_{m0}$.
To determine the constraints on $\phi_i$ and $\Omega_{m0}$ we  vary $\phi_i$ in the
range $[0.9, 1.5]$ and $\Omega_{m0}$ in the range $[0.2, 0.4]$. 
Fig.\ref{chifixq} shows the $1\sigma $ and $2\sigma$ contours in
the $(\phi_i, \Omega_{m0})$ parameter space for different values of $\beta$
keeping $q_{\mathcal{M}}$ fixed to $0.6$. We see that for higher values of $\beta$, the lower values 
of $\phi_i$ and higher values of $\Omega_{m0}$ is allowed.
Fig.\ref{chifixbeta} shows the contours in parameter space
 $(\phi_i, \Omega_{m0})$, for different values of $q_{\mathcal{M}}$
fixing $\beta=0.8$. We see that for higher values of $q_{\mathcal{M}}$, the parameter space gets
more and more constrained, allowing a very small range of initial values of field
$\phi_i$.

The $\chi^2 $ statistical measures described above are based on the assumption 
that the underlying model is a viable one. The analysis is good  at  finding  the
best parameters in a model that will fit the data,
but is insufficient for deciding whether the model itself is the best one.  
To assess the strength of the models one uses the 
information criteria (IC)~\cite{ic}. The statistics favour models that give a good fit
with fewer parameters. We use the Bayesian information
criterion (BIC) and Akaike information criterion (AIC) to
select the best fit models in comparison to the standard $\Lambda$CDM model.
The AIC and BIC are defined as
\begin{equation}
 AIC =-2 ln L+ 2k\,,
\end{equation}
\begin{equation}
 BIC =-2 ln L+ klnN\,,
\end{equation}
where $L$, $k$ and $N$ are the maximum likelihood, the number of parameters,
and  the number of data points, respectively. For
Gaussian errors, $\chi^2=-2 ln L$.  In order to compare it to the $\Lambda$CDM model
we find the difference 
$\Delta$AIC and $\Delta$BIC from the $\Lambda$CDM model of the best fit values 
of the present model for various $q_{\mathcal{M}}$ and $\beta$ values. In table
 $\ref{aic}$, $\Delta$AIC and $\Delta$BIC are shown, from which we can
 infer the strength of the model fixed by particular $q_{\mathcal{M}}$ and $\beta$ in comparison
 to the $\Lambda$CDM model. From the table we notice that models for greater
 $q_{\mathcal{M}}$ and $\beta$ are poorly fitted by the data. For small values of $q_{\mathcal{M}}$ and $\beta$ the model is better fitted than
 some of the models of dark energy, for instance the Dvali-Gabadadze-Poratti 
 braneworld model~\cite{DGP}, considered in Ref.~\cite{ic}.

\begin{table*}\centering
\begin{tabular}{|c|c|c|c|}
\hline
 $q_{\mathcal{M}}$ & $\beta$  &$\Delta$AIC &$\Delta$BIC\\
\hline\hline
 $.6$ &  $.6$ & $8$ & $13$  \\
\hline
$.6$ &  $.7$ & $9$ & $13$  \\
\hline
$.6$ &  $.8$ & $10$ & $14$  \\
\hline
$.6$ &  $.9$ & $10$ & $15$  \\
\hline
$.7$ &  $.8$ & $24$ & $28$  \\
\hline
$.8$ &  $.8$ & $26$ & $31$  \\
\hline
$.9$ &  $.8$ & $28$ & $33$  \\
\hline
\end{tabular}
\caption{Values of $\Delta$AIC and $\Delta$BIC for different values of $q_{\mathcal{M}}$ and $\beta$.}
\label{aic}
\end{table*}

\section{Conclusion}

In this paper we have employed the Buchert formalism~\cite{Buchert:1999er, buchert2, buchert3, buchert4}
 for obtaining the averaged
effect of inhomogeneities on the late evolution of the universe.
The cosmological evolution is studied here in context of a model constructed using a two-scale void-wall partitioning~\cite{weigand,bose}. We have
 performed a detailed analysis of the various
aspects of the future evolution of the presently accelerating universe 
within this framework.  The equations governing the dynamics at large global scales in this formalism suggest themselves to interpret the sources of inhomogeneities in geometrical variables
in terms of an effective scalar field~\cite{morf1}. We thus choose to consider the model with averaged
inhomogeneities at large scales as a variant of the standard FLRW model with matter evolving in a
mean field of  scalar field $\phi$ representing the backreaction terms.
The potential of this scalar field captures the geometrical  signatures
of structures in universe and is free from phenomenological parametrisation. The investigation of the field dynamics in the
global domain, where the the cosmological principle is intact is done in details. It is seen that the field resembling a thawing dark energy model gives rise
to a transient acceleration at present. 

We have  imposed observational constraints on the
model parameters $\phi_i$ and $\Omega_{m0}$ using the data from SNIa~\cite{SnIa}, 
BAO~\cite{bao1,bao2,bao3},
and CMB~\cite{cmb} observations. By constructing the corresponding
contour plots we deduce that $\phi_i$ is constrained by the
data to a small range of values, depending upon the values of individual expansion
rates of the two sub-domains, {\it i.e.}, the void and the wall.
The present density parameter of matter $\Omega_{m0}$ is constrained around the concordance values. We have also checked the strength of the 
model in comparison with the $\Lambda$CDM model using Bayesian information
criterion (BIC) and Akaike information criterion (AIC)~\cite{ic}. Though the present toy model is not favoured by the data, it still enables a better fit compared to some other
proposed models of dark energy involving modifications~\cite{DGP} of standard GR. 
The results obtained by us in the present
 work follow from the choice of our ansatz regarding the wall and
 void scale factors which reflect, at best, the qualitative expectations of the physical  behaviour of overdense and underdense regions, respectively. Further work remains to be done where more
 realistic and observationally compatible properties of the void
 and wall regions could
 be incorporated within the formalism.
Further investigations in context of  
more realistic models involving multidomain scenarios is called for, especially
in view of a remarkable feature of our results 
showing that the averaged effect of inhomogeneities
render the current global acceleration to a transient phenomenon, which dies down
in future.

\section{ACKNOWLEDGEMENTS}
The authors thank Thomas Buchert and Nilok Bose for useful discussions.

\end{document}